\documentclass{article}
\usepackage{natbib,epsfig}

\usepackage{amsthm,amsmath,amssymb,colonequals,multirow,graphicx,epstopdf}

\newcommand{\vecx}{\mathbf{x}}
\newcommand{\vecX}{\mathbf{X}}

\newcommand{\vecz}{\mathbf{z}}
\newcommand{\vectau}{\mbox{\boldmath$\tau$}}
\newcommand{\vecZ}{\mathbf{Z}}

\newcommand{\Ew}{\mathbb{E}\left[ W_i \mid \vecx_i \right]}
\newcommand{\Ewinv }{\mathbb{E}\left[1/W_i \mid \vecx_i \right]}
\newcommand{\Ewjnv }{\mathbb{E}\left[1/W_j \mid \vecx_j \right]}
\newcommand{\Elogw}{\mathbb{E}\left[ \log{W_i} \mid \vecx_i \right]}

\newcommand{\varthet}{\mbox{\boldmath$\vartheta$}}
\newcommand{\vectheta}{\mbox{\boldmath$\theta$}}
\newcommand{\vecmu}{\mbox{\boldmath$\mu$}}

\newcommand{\vecalpha}{\mbox{\boldmath$\alpha$}}
\newcommand{\vecpi}{\mbox{\boldmath$\pi$}}
\newcommand{\mSigma}{\mbox{\boldmath$\Sigma$}}
\newcommand{\matsig}{\mSigma}

\newcommand{\lprod}{\prod_{i=1}^n}
\newcommand{\lsum}{\sum_{i=1}^n}
\newcommand{\lssum}{\sum_{j=1}^n}
\newcommand{\gsum}{\sum_{g=1}^G}
\newcommand{\gprod}{\prod_{g=1}^G}
\newcommand{\zig}{\hat{\tau}_{ig}}
\newcommand{\zjg}{\hat{\tau}_{jg}}

\begin{document}

\title{Mixtures of Shifted Asymmetric Laplace Distributions}

\author{Brian~C.~Franczak,
        Ryan~P.~Browne,
        and~Paul~D.~McNicholas\thanks{Department of Mathematics \& Statistics, University of Guelph, Guelph, Ontario, N1G 2W1, Canada. E-mail: paul.mcnicholas@uoguelph.ca.}
}
\date{Department of Mathematics \& Statistics, University of Guelph.}
\maketitle

\begin{abstract}
A mixture of shifted asymmetric Laplace distributions is introduced and used for clustering and classification. A variant of the EM algorithm is developed for parameter estimation by exploiting the relationship with the general inverse Gaussian distribution. This approach is mathematically elegant and relatively computationally straightforward. Our novel mixture modelling approach is demonstrated on both simulated and real data to illustrate clustering and classification applications. In these analyses, our mixture of shifted asymmetric Laplace distributions performs favourably when compared to the popular Gaussian approach. This work, which marks an important step in the non-Gaussian model-based clustering and classification direction, concludes with discussion as well as suggestions for future work.
\end{abstract}

\section{Introduction}

Finite mixture models are based on the underlying assumption that a population is a convex combination of a finite number of densities. They therefore lend themselves quite naturally to classification and clustering problems. Formally, a random vector $\mathbf{X}$ arises from a parametric finite mixture distribution if, for all $\vecx \subset \mathbf{X}$, we can write its density as 
$$f(\vecx\mid\varthet)= \sum_{g=1}^G \pi_g f_g(\vecx\mid\vectheta_g),$$ where $\pi_g >0$, such that $\sum_{g=1}^G \pi_g = 1$ are the mixing proportions, $f_1(\vecx\mid\vectheta_g),\ldots,f_G(\vecx\mid\vectheta_g)$ are called component densities, and $\varthet=(\vecpi,\vectheta_1,\ldots,\vectheta_G)$ is the vector of parameters with $\vecpi=(\pi_1,\ldots,\pi_G)$. 
The component densities $f_1(\vecx\mid\vectheta_1),\ldots,f_G(\vecx\mid\vectheta_G)$ are usually taken to be of the same type, most often multivariate Gaussian. In the event that the component densities are multivariate Gaussian, the density of the mixture model is  $f(\vecx\mid\varthet)= \sum_{g=1}^G \pi_g\phi(\vecx\mid\vecmu_g,\matsig_g)$, where $\phi(\vecx\mid\vecmu_g,\matsig_g)$ is the multivariate Gaussian density with mean~$\vecmu_g$ and covariance matrix $\matsig_g$. The popularity of the multivariate Gaussian distribution is due to its mathematical tractability and its flexibility in capturing densities; we will return to this latter point in Section~\ref{sec:XXX}. Herein, we shall follow convention and use the term model-based clustering to mean clustering using mixture models. Model-based classification \citep[e.g.,][]{mcnicholas10c}, or partial classification \citep[cf.][Section~2.7]{mclachlan92}, can be regarded as a semi-supervised version of model-based clustering. 

At the time of the review paper of \cite{fraley02a}, almost all work on clustering and classification using mixture models had been based on Gaussian mixture models (e.g.,  \cite{banfield93}, \cite{celeux95}, \cite{ghahramani97}, \cite{fraley98}, \cite{tipping99b}, \cite{mclachlan00b}, and \cite{mclachlan00a}, amongst others). An important example of non-Gaussian work from this time is the early work on clustering using mixtures of multivariate $t$-distributions carried out by \cite{mclachlan98} and \cite{peel00}. This work was the forerunner to several papers on clustering using mixtures of multivariate $t$-distributions, including those by \cite{mclachlan07}, \cite{andrews11b}, and \cite{baek11}.  Work has also burgeoned on skew-normal distributions \citep[e.g.,][]{lin09}, skew-$t$ distributions \citep[e.g.,][]{lin10,lee11,vrbik12}, and other non-elliptically contoured distributions \citep[e.g.,][]{karlis07,karlis09,browne11}. 


The recent burgeoning of non-Gaussian approaches to model-based clustering and classification has coincided with yet more papers on Gaussian approaches. These include work on extensions of mixtures of factor analyzers \citep{mcnicholas08,mcnicholas10d,baek10} and developments on variable selection and dimension reduction for Gaussian model-based clustering \citep{raftery06,maugis09,scrucca09}. Nevertheless, the fecundity of non-Gaussian approaches has certainly been more potent than that of Gaussian approaches over the last few years. This paper introduces a non-Gaussian approach that allows for skewness while also parameterizing location and scale. Our approach is effective while also being mathematically elegant and relatively computationally straightforward. The methodology is developed in Section~\ref{sec:method}, parameter estimation via deterministic annealing and an expectation-maximization algorithm is outlined in Section~\ref{EM}, and both simulated and real data analyses are used to illustrate our approach (Section~\ref{sec:data}). The paper concludes with a summary and suggestions for future work (Section~\ref{sec:summ}).

\section{Methodology}\label{sec:method}

\subsection{Generalized Inverse Gaussian Distribution}
The density of a random variable $X$ following a generalized inverse Gaussian (GIG) distribution is given by 
\begin{equation}\label{gig}
q(x) = \frac{(a/b)^{\nu/2}x^{\nu-1}}{2K_{\nu}(\sqrt{ab})} \exp\left\{-\frac{ax+b/x}{2}\right\},
\end{equation}
for $x>0$, where $a,b\in\mathbb{R}^+$, $\nu\in\mathbb{R}$, and $K_{\nu}$ is the modified Bessel function of the third kind with index~$\nu$. There are several special cases of the GIG distribution, such as the gamma distribution ($b=0$, $\nu>0$) and the inverse Gaussian distribution ($\nu=-1/2$). The GIG distribution was introduced by \cite{good53} and its statistical properties were laid down by \cite{barndorff77}, \cite{blaesild78}, \cite{halgreen79}, and \cite{jorgensen82}. It has some attractive properties including the tractability of the following expected values: 
\begin{equation}\begin{split}\label{eqn:exp_vals}
&\mathbb{E}\left[ X \right] =
\frac{\sqrt{b}K_{\nu+1}\left( \sqrt{ab}\right) }{\sqrt{a}K_{\nu}\left( \sqrt{ab}\right)},\qquad\mathbb{E}\left[{1}/{X}\right] = 
\frac{\sqrt{a}K_{\nu+1}\left( \sqrt{ab}\right)}{\sqrt{b}K_{\nu}\left( \sqrt{ab}\right)} -\frac{2\nu}{b}.
\end{split}\end{equation}

\subsection{Shifted Asymmetric Laplace Distribution}\label{intro}
Consider a $p$-dimensional random vector $\mathbf{Z}$ from a centralized asymmetric Laplace (CAL) distribution \citep{kotz01}. The density of $\mathbf{Z}$ is given by
\begin{equation}\begin{split}\label{CAL}
f(\vecz\mid\vecalpha, \mSigma)=
\frac{2K_{\nu}\left(u\right)}
{(2\pi)^{p/2}\vert\mSigma\vert^{1/2}}&
\left(\frac{\vecz'\mSigma^{-1}\vecz}
{2+\vecalpha'\mSigma^{-1}\vecalpha}\right)^{\nu/2}\exp\{\vecz'\mSigma^{-1}\vecalpha\},
\end{split}\end{equation}
where $\nu=(2-p)/2$, $u = \sqrt{(2+\vecalpha'\mSigma^{-1}\vecalpha)\left(\vecz'\mSigma^{-1}\vecz\right)}$, 
$\mSigma$ is a covariance matrix, and $\vecalpha\in\mathbb{R}^p$ represents the skewness in each dimension. \cite{kotz01} use the notation $\vecZ \backsim \mathcal{A L}_p \left( \vecalpha, \mSigma\right)$ to indicate that the random variable $\vecZ$ follows a $p$-dimensional CAL distribution and provide an extensive properties list.

The CAL density \eqref{CAL} is prohibitive for model-based clustering and classification applications because it would force each component density to be joined at the same origin. To address this problem, consider $\vecZ \backsim \mathcal{A L}_p \left( \vecalpha, \matsig\right)$ and introduce a shift parameter $\vecmu\in\mathbb{R}^p$ by considering a random vector $\vecX = (\vecZ+\vecmu)\backsim \mathcal{SAL}_p(\vecalpha,\matsig,\vecmu)$, where $\mathcal{SAL}_p(\vecalpha,\matsig,\vecmu)$ denotes a $p$-dimensional shifted (non-centralized) asymmetric Laplace (SAL) distribution with density given by
\begin{equation}\begin{split}\label{SAL}
\xi\left(\vecx\mid\vecalpha, \matsig,\vecmu\right)&=
\frac{2\exp\{(\vecx-\vecmu)'\matsig^{-1}\vecalpha\}}
{(2\pi)^{p/2}\vert\matsig\vert^{1/2}}\left(\frac{\delta\left(\vecx,\vecmu\mid\matsig\right)}
{2+\vecalpha'\matsig^{-1}\vecalpha}\right)^{\nu/2}
K_{\nu}\left(u\right),
\end{split}\end{equation}
where $u = \sqrt{(2+\vecalpha'\matsig^{-1}\vecalpha)\delta\left(\vecx,\vecmu\mid\matsig\right)}$, $\delta\left(\vecx,\vecmu\mid\matsig\right)=\left(\vecx-\vecmu\right)'\matsig\left(\vecx-\vecmu\right)$ is the squared Mahalanobis distance between $\vecx$ and $\vecmu$, and $\nu$, $\vecalpha$, and $\matsig$ are defined as before.

\cite{kotz01} note that the random variable $\vecZ\backsim\mathcal{AL}_p\left(\vecalpha,\matsig\right)$ can be generated through the relationship $\vecZ=W\vecalpha+\sqrt{W}\mathbf{Y}$, where $W$ is a random variable from an exponential distribution with mean~1 and $\mathbf{Y}\backsim\mathcal{N}(\mathbf{0},\matsig)$ is generated independent of $W$. Therefore, the random variable $\vecX\backsim\mathcal{S A L}_p(\vecalpha,\matsig,\vecmu)$ can be generated through the relationship $\vecX=\vecmu+W\vecalpha+\sqrt{W}\mathbf{Y}$ and so $\vecX\mid{W=w}\backsim\mathcal{N}\left(\vecmu+w\vecalpha,w\matsig\right)$.

The distribution of $W$ conditional on the data can be computed through the use of Bayes' theorem, i.e.,
$$f_W(w\mid\vecX =\vecx) = {f_\vecX(\vecx\mid W=w)h(w)}/{f_\vecX(\vecx)},$$ 
where $\vecX\mid W=w\backsim\mathcal{N}(\vecmu+w\vecalpha,w\matsig)$,
$W\backsim\exp(1)$, and $f_\vecX(\vecx)$ is the density of the shifted asymmetric Laplace distribution given in~\eqref{SAL}. It follows that
\begin{equation}\begin{split}\label{densw}
f_W(w&\mid\vecX =\vecx)
=\frac{w^{\nu-1}}{2}
\left(\frac{\delta\left(\vecx,\vecmu\mid\matsig\right)}{2+\vecalpha'\matsig^{-1}\vecalpha}\right)^{-\nu/2}\frac{\exp\left\{-\frac{1}{2w}\delta\left(\vecx,\vecmu\mid\matsig\right)-\frac{w}{2}\left(2+\vecalpha'\matsig^{-1}\vecalpha\right)\right\}}{K_{\nu}\left(\sqrt{(2+\vecalpha'\matsig^{-1}\vecalpha)\delta(\vecx,\vecmu\mid\matsig)}\right)},
\end{split}\end{equation}
where $\nu$, $\vecalpha$, $\vecmu$, $\matsig$, and $\delta\left(\vecx,\vecmu\mid\matsig\right)$ are as defined for~\eqref{SAL}. Recalling the density of a GIG random variable \eqref{gig}, it then follows from \eqref{densw} that $f_W(w\mid\vecX =\vecx)$ is the GIG density with $a\equiv2+\vecalpha'\matsig^{-1}\vecalpha$ and $b\equiv\delta(\vecx,\vecmu\mid\matsig)$, cf.\ \citep{barndorff97}.

We introduce a finite mixture of SAL distributions so that the $g$th component density is $\mathcal{SAL}_p(\vecalpha_g,\matsig_g,\vecmu_g)$, where the parameters are as defined for \eqref{SAL}. The density of a mixture of SAL distributions is 
$f(\mathbf{x}\mid\varthet)=\gsum{\pi_g\xi(\vecx\mid\vecalpha_g, \matsig_g,\vecmu_g)}$,
where $\varthet$ is the vector of all model parameters and $\xi(\vecx\mid\vecalpha_g, \matsig_g,\vecmu_g)$ is the SAL density from \eqref{SAL}. 

\section{Parameter Estimation}\label{EM}
\subsection{EM Algorithm}\label{sec:mbclust}
The expectation-maximization (EM) algorithm \citep{dempster77} is an iterative procedure for finding the maximum likelihood estimates when data are incomplete or are treated as such. EM algorithm computations are based on the complete-data likelihood, i.e., the likelihood of the observed data plus the latent or missing data. In the E-step, the expected value of the complete-data log-likelihood is computed, and in the M-step, this value is maximized with respect to the model parameters. The E- and M-steps are then iterated until some convergence criterion is attained.

A common criticism of the use of the EM algorithm for model-based clustering is that the singularity-riddled likelihood surface makes parameter estimation unreliable and heavily dependent on the starting values. To help overcome this problem, \cite{zhou09} introduced a deterministic annealing algorithm that flattens the likelihood surface by introducing an auxiliary variable $v\in [0,1]$. We will illustrate this approach by using a version of this deterministic annealing algorithm in conjunction with an EM algorithm to fit our SAL mixture model (cf.\ Section~\ref{sec:imp}). Note that \cite{eltoft06} give an example of an EM-type algorithm for fitting a multivariate Laplace distribution.

\subsection{Application to Mixture of SAL Distributions}
For our SAL mixture models, the complete-data comprise the observed $\vecx_1,\ldots,\vecx_n$, the component membership labels $\vectau_1,\ldots\vectau_n$, and the variable $W$. For each $i$, we have $\vectau_i=(\tau_{i1},\ldots,\tau_{iG})$, where $\tau_{ig}=1$ if observation~$i$ is in component~$g$ and $\tau_{ig}=0$ otherwise, for $i=1,\ldots,n$ and $g=1,\ldots,G$. 
The complete-data likelihood is given by
\begin{equation}\label{CDL}
\mathcal{L}
=\lprod\gprod{\lbrack\pi_g\phi\left(\vecx_i\mid\vecmu_g+w_i\vecalpha_g,w_i\mSigma_g\right)h\left(w_i\right)\rbrack}^{\tau_{ig}},
\end{equation}
with the same notation used previously and where $\phi\left(\vecx_i\mid\vecmu_g+w_i\vecalpha_g,w_i\mSigma_g\right)$ is the density of a multivariate Gaussian distribution with mean $\vecmu_g+w_i\vecalpha_g$ and covariance matrix $w_i\mSigma_g$. 
The expected-value of the complete-data log-likelihood is given by
\begin{equation*}\begin{split}
&\mathcal{Q} = \gsum n_g\log{\pi_g}-\frac{np}{2}\log{2\pi} - \frac{np}{2}\lsum{\Elogw} + \gsum\frac{n_g}{2}\log{\left| \mSigma_g^{-1}\right|} + 2\lsum\gsum\zig\left(\vecx_i-\vecmu_g\right)'\mSigma_g^{-1}\vecalpha_g\\
&- \frac{1}{2}\lsum\gsum\zig  \left( \vecx_i - \vecmu_g\right)'\Ewinv\mSigma_g^{-1}\left(\vecx_i-\vecmu_g\right)
- \frac{1}{2}\lsum\gsum\zig \Ew\vecalpha_g'\mSigma_g^{-1}\vecalpha_g\ - \lsum\gsum\zig\Ew,
\end{split}\end{equation*}
where $n_g = \lsum \zig$ and 
\begin{equation}
\zig\colonequals\mathbb{E}\left[\tau_{ig}\mid\vecx_i\right]=\frac{\pi_g\xi\left(\vecx_i\mid\vecalpha_g, \matsig_g,\vecmu_g\right)}{\sum_{j=1}^G\pi_j\xi\left(\vecx_i\mid\vecalpha_j, \matsig_j,\vecmu_j\right)}.
\end{equation}
%
The expected values $\Ew$ and $\Ewinv$ are computed using the formulae in \eqref{eqn:exp_vals}.

In the M-step, we maximize $\mathcal{Q}$ with respect to the model parameters to get the updates. Specifically, the mixing proportions, skewness parameter, and shift parameter are updated by 
$\hat{\pi}_g = {n_g}/{n}$,
\begin{equation*}\begin{split}
\hat{\vecalpha}_g&=\frac{\lsum \zig\Ewinv\lssum\zjg\vecx_j-n_g\lsum \zig\Ewinv\vecx_i}{\lsum \zig\Ew\lssum\zjg\Ewjnv - n_g^2},
\end{split}\end{equation*}
and
\begin{equation*}\begin{split}
\hat{\vecmu}_g &= \frac{\lsum\zig\Ew\lssum\Ewjnv\vecx_j- n_g\lsum\zig\vecx_i}{\lsum \zig\Ew\lssum\Ewjnv-n_g^2},
\end{split}\end{equation*}
respectively. Each component covariance matrix $\hat{\mSigma}_g$ is updated by
\begin{equation}\label{mSigma}
\hat{\mSigma}_g = 
\mathbf{S}_g - \hat{\vecalpha}_g\mathbf{r}_g' 
- \mathbf{r}_g\hat{\vecalpha}_g'
+ \frac{1}{n_g}\hat{\vecalpha}_g\hat{\vecalpha}_g'\lsum\zig\Ew,
\end{equation} 
where $\mathbf{S}_g=({1}/{n_g})\lsum\zig\Ewinv\left(\vecx_i-\hat{\vecmu}_g\right)\left(\vecx_i-\hat{\vecmu}_g\right)'$ and $\mathbf{r}_g = ({1}/{n_g})\lsum\zig\left(\vecx_i-\hat{\vecmu}_g\right)$.

The E- and M-steps are iterated until convergence. Practically, the E-step consists of updating the values of the expected values $\zig$, $\Ew$, and $\Ewinv$. At the first iteration, these updates are based on the initialized values of the parameter estimates $\hat{\pi}_g$, $\hat{\vecalpha}_g$, $\hat{\vecmu}_g$, and $\hat{\mSigma}_g$ (cf.\ Section~\ref{sec:imp}). For all other iterations, these updates are based on the values of the parameter estimates from the previous iteration. The M-step consists of updating the values of the parameters $\hat{\pi}_g$, $\hat{\vecalpha}_g$, $\hat{\vecmu}_g$, and $\hat{\mSigma}_g$ based on the expected values from the E-step. The E- and M-steps are iterated until convergence (cf.\ Section~\ref{sec:converg}).

At convergence, the $\zig$ are the \textit{a posteriori} probabilities of component membership for each observation and can be used to cluster the observations into groups. Predicted classifications are obtained via maximum \textit{a posteriori} (MAP) probabilities, where $\text{MAP}\{\zig\}=1$ if max$_g\{\zig\}$ occurs at component $g$ and $\text{MAP}\{\zig\}=0$ otherwise.

\subsection{Initialization}\label{sec:imp}

The deterministic annealing algorithm \citep{zhou09} is the same as the EM algorithm described in Section~\ref{sec:mbclust}, except that now
\begin{equation}\label{zigv}
\mathbb{E}\left[\tau_{ig}\mid\vecx_i\right]=\frac{[\pi_g\xi\left(\vecx_i\mid\vecalpha_g, \matsig_g,\vecmu_g\right)]^v}{\sum_{h=1}^G[\pi_h\xi\left(\vecx_i\mid\vecalpha_h, \matsig_h,\vecmu_h\right)]^v}
\end{equation}
in each E-step. Here, the auxiliary parameter $v\in[0,1]$, which is drawn from an increasing sequence of user-specified length, transforms the likelihood surface to improve the chances of finding the dominant mode. The user-specified sequence runs from $0$ to $1$ and its length determines how many iterations of the deterministic annealing algorithm will be preformed.
The annealing algorithm itself is initialized using random starting values of $\pi_g$, $\vecalpha_g$, $\mSigma_g$, and $\vecmu_g$. 
In our analyses (Section~\ref{sec:data}), we run the deterministic annealing algorithm ten times and choose the values that give the highest likelihood as the starting values for our EM algorithms.

\subsection{Convergence}\label{sec:converg}

\subsubsection{Aitken acceleration}
The Aitken acceleration \citep{aitken26} is used to determine convergence of our EM algorithms. An EM algorithm can be considered to have converged when
$l_{\infty}^{(k+1)} - l^{(k+1)}<\epsilon$,
where $l^{(k+1)}$ is the log-likelihood at iteration $k+1$ and 
\begin{equation*}
l_{\infty}^{(k+1)} = l^{(k)} + \frac{l^{(k+1)}-l^{(k)}}{1 - a^{(k)}}
\end{equation*}
is an asymptotic estimate of the log-likelihood at iteration $k+1$ \citep[cf.][]{bohning94}.
The value
$a^{(k)}={[l^{(k+1)} - l^{(k)}]}/{[l^{(k)} - l^{(k-1)}]}$
is the Aitken acceleration at iteration~$k$. For the analyses herein, we use a slightly modified convergence criterion, stopping our EM algorithms when
$l_{\infty}^{(k+1)} - l^{(k)}<\epsilon$ \cite[cf.][]{lindsay95}. 

\subsubsection{Dealing with infinite likelihood}\label{estimatemu}
As our EM algorithm iterates, we must handle the complications that arise when computing~$\hat{\vecmu}_g$. Specifically, as the algorithm iterates towards convergence, the value of $\hat{\vecmu}_g$ will tend to an observation~$\vecx_i$. This happens because of the $[\delta(\vecx,\vecmu\mid\mSigma)/(2+\vecalpha'\mSigma^{-1}\vecalpha)]^{\nu/2}$ term in the multivariate SAL density~\eqref{SAL}. Although these estimates of $\hat{\vecmu}_g$ maximize the likelihood, they create computational issues when trying to determine the remaining parameter values and, specifically, the expected value $\Ewinv$. 

To overcome this problem, we stop searching for $\vecmu_g$ when $\hat{\vecmu}_g$ has the same value as some $\vecx_i$ because the log-likelihood becomes infinite at this point.
We proceed by taking the value of $\hat{\vecmu}_g$ at the iteration before it becomes equal to any $\vecx_i$ (we denote this value $\hat{\vecmu}^*_g$) as the estimate for $\vecmu_g$. We then update of $\vecalpha_g$ using
\begin{equation*}
\hat{\vecalpha}_g^*=\left[\frac{\lsum\zig\left(\vecx_i-\vecmu_g^*\right)'}{\lsum\zig \Ew}\right]',
\end{equation*}
and update $\mSigma_g$ given the update in (11).
We use a different approach to overcome this problem in the deterministic annealing algorithm, simply restricting the expected value of $\mathbb{E}[1/W_i \mid \vecx_i]$ from exceeding a value of $-\log (1 - v)$ at each iteration. We acknowledge that our solution to this problem is a simple-minded one. However, we have found it to be quite effective and a more thorough exploration of this problem in general is the subject of ongoing work. 

\subsection{Model-Based Classification}
Suppose we have $n$ observations and $k$ of these observations have known group memberships. We can use the group memberships of these $k$ observations to estimate memberships for the remaining $n-k$ observations within a joint likelihood framework. This approach, known as model-based classification, is a semi-supervised version of model-based clustering. Without loss of generality, we order the observations $\vecx_1,\ldots,\vecx_k,\vecx_{k+1},\ldots,\vecx_n$ so that the first~$k$ have known group memberships. Therefore, the values of $\tau_{ig}$ are known for $i=1,\ldots,k$ and the SAL model-based classification likelihood is given by 
\begin{equation}\label{joint}\begin{split}
\mathcal{L}&\left(\vecx_1,\ldots,\vecx_n,\vectau_1,\ldots,\vectau_n\mid\varthet\right)=
\prod_{i=1}^k\prod_{g=1}^G{\lbrack\pi_g\xi\left(\vecx_i\mid\vecalpha_g, \matsig_g,\vecmu_g\right)\rbrack^{\tau_{ig}}}\prod_{j=k+1}^{n}\sum_{h=1}^H{\pi_h\xi\left(\vecx_j\mid\vecalpha_h, \matsig_h,\vecmu_h\right)},
\end{split}\end{equation}
where $H\geq G$. Parameter estimation is carried out in an analogous fashion to model-based clustering.

\subsection{Model Selection and Performance}

The Bayesian information criterion \citep[BIC;][]{schwarz78} is commonly used for Gaussian mixture model selection: 
$$\text{BIC}=2l(\vecx\mid\hat{\varthet})-\nu\log{n},$$
where $l(\vecx\mid\hat{\varthet})$ is the maximized log-likelihood, $\hat{\varthet}$ is the maximum likelihood estimate of $\varthet$, $\mathtt{\nu}$ is the number of free parameters in the model, and $n$ is the number of observations. In the analyses herein, we prefer to use a model selection approach that is specifically designed for clustering and classification applications. To this end, we use the integrated classification likelihood \citep[ICL;][]{biernacki00}, which is calculated using
$$\text{ICL} \approx \text{BIC} + \sum_{i=k+1}^n\sum_{g=1}^G{\text{MAP}\{\hat{z}_{ig}\}\log{\hat{z}_{ig}}},$$
where $\sum_{i=k+1}^n\sum_{g=1}^G{\text{MAP}\{\hat{z}_{ig}\}\log{\hat{z}_{ig}}}$, the estimated mean entropy, reflects the uncertainty in the classification of observation $i$ into group $g$. 

To assess clustering and classification performance, we use a cross tabulation of our MAP classifications against the true group memberships. We then compute the adjusted Rand index \citep[ARI;][]{hubert85}. The Rand index \citep{rand71} was introduced to compare partitions. It is the ratio of pairs that should be and are together plus pairs that should be and are apart, divided by the total number of pairs. The Rand index takes a value between 0 and 1, where 1 indicates perfect agreement. An unattractive feature of the Rand index is that it has a positive expected value under random classification. To correct this undesirable property, \cite{hubert85} introduced the ARI to account for chance agreement. The ARI also takes a value of 1 when classification is perfect but has an expected value of 0 under random classification. The ARI can also take negative values and this happens for classifications that are worse than would be expected by chance.

\section{Data Analyses}\label{sec:data}
\subsection{Introduction}\label{sec:XXX}
The SAL mixture models are applied to simulated data (Section~\ref{sec:SimuStudy1}) and to two real data sets: the famous Old Faithful geyser data (Section~\ref{sec:OldFaith}) and data on cellular localization sites for proteins in yeast (Section~\ref{sec:Yeast}). The simulation study was conducted to assess the accuracy of our SAL parameter estimates, including selection of the number of components, and to draw comparison with the Gaussian mixture model approach. In the real analyses, we illustrate the SAL approach, judging its performance alongside that of Gaussian mixtures. 

On the face of it, a comparison of our SAL approach to Gaussian mixtures might be considered a little empty because one would expect Gaussian mixtures to use more than one component to model a cluster with skewness. Indeed, there has been a lot of work within the literature on merging Gaussian components \citep[e.g.,][]{baudry10,hennig10}. However, our examples illustrate that when predicted classifications differ for the SAL and Gaussian approaches, merging Gaussian components cannot always be used to rectify shortcomings in the Gaussian results (cf. Sections~\ref{sec:SimuStudy1} and~\ref{sec:Yeast}). One may also argue that the mixture of multivariate skew-$t$ distributions could also be used to accommodate skewness. This is true, but two different representations of the mixture of skew-$t$ distributions have appeared within the literature \citep[cf.][]{sahu03,pyne09} and neither have the elegance of SAL mixtures; this is most apparent in the intractability of the E-step for the skew-$t$ approach \citep[cf.][]{lin10,lee11,vrbik12}.

Note that, for all analyses, we use the same deterministic annealing starting values for the mixtures of Gaussian distributions as for our SAL mixtures. Further, we treat each analysis as a genuine clustering problem by removing all labels; we then apply SAL and Gaussian mixture models.

\subsection{Simulation Study}\label{sec:SimuStudy1}
\begin{figure}[!h]
\centering%
\includegraphics[width=3.45in,height=2.45in]{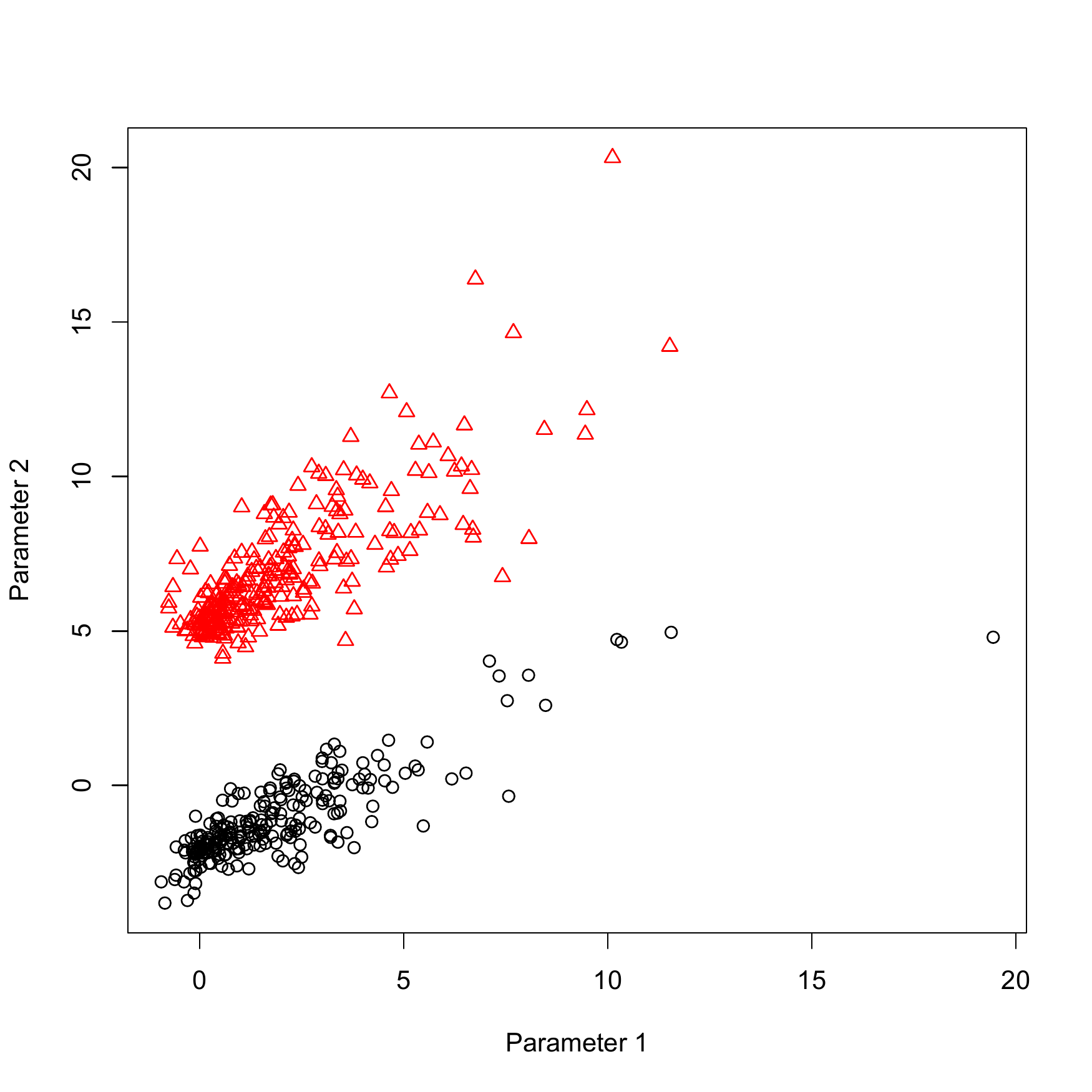}
\vspace{-0.1in}
\caption{Example of a simulated data set ($n = 500$), coloured by component.\label{SimuData1}}
\end{figure}

We used the relationship between the SAL and Gaussian distributions (cf. Section~\ref{intro}) to generate multivariate SAL data. Specifically, we simulated 25 data sets for $n=500$ with $p = 2$ dimensions and $G = 2$ components (e.g., Figure~\ref{SimuData1}). The data were generated using skewness parameters $\vecalpha_1 = (2,1)$ and $\vecalpha_2 = (2,2)$, shift parameters $\vecmu_1 = (0,-2)$ and $\vecmu_2 = (0,5)$, and covariance matrices $$\mSigma_1 = \begin{pmatrix}1 & 0.5\\ 0.5 & 1\end{pmatrix} \qquad \text{and} \qquad \mSigma_2 =  \begin{pmatrix}1 & 0\\ 0 & 1\end{pmatrix}.$$ 

\begin{figure}[htb!]
\centering%
\includegraphics[width=3.45in,height=2.45in]{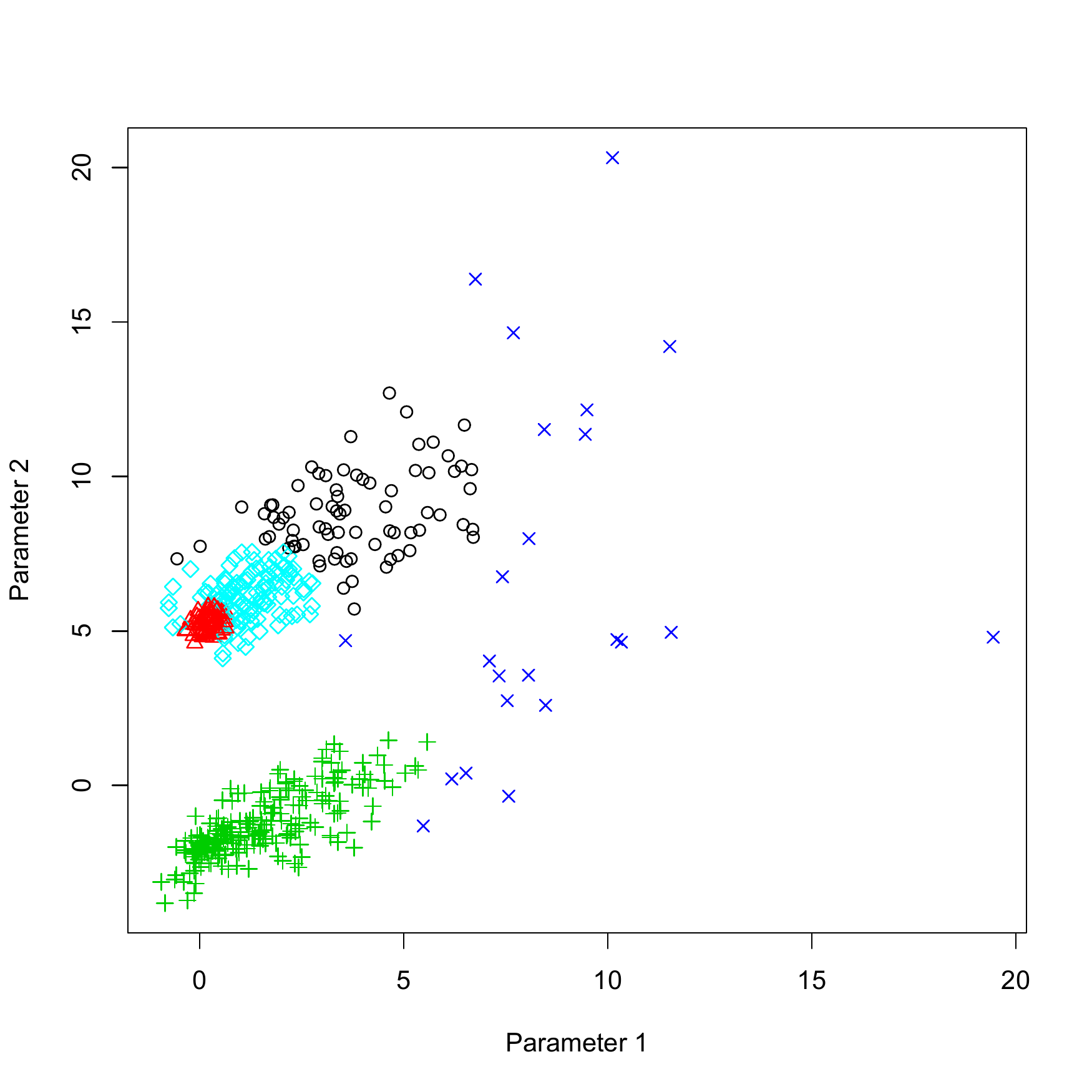}
\vspace{-0.1in}
\caption{One of the $G=5$ Gaussian solutions for the simulated data, coloured by MAP classification results.\label{fig:gaussEasy}}
\end{figure}

The clustering results (Table~\ref{SimuClust2}) show that the SAL mixtures gave almost perfect clustering performance over all 25 runs (average ARI=$0.9968$). The Gaussian approach, however, gave relatively poor clustering performance (average ARI=$0.4988$) and never returned the correct number of clusters. The most common Gaussian solution (chosen 21 times) has five components, e.g., Figure~\ref{fig:gaussEasy}, and each of the other four solutions had four components.   
\begin{table}[ht]
\centering
\caption{Summary for the analysis of the 25 simulated data sets using our SAL mixtures, and using the Gaussian approach.\label{SimuClust2}}
\vspace{.1in}
\begin{tabular*}{0.48\textwidth}{@{\extracolsep{\fill}}ll|r}
\hline
\multirow{2}{*}{SAL}&$G=2$ selected & 100\% \\
&Average ARI (std.\ dev.) & $0.9968$ ($0.00516$)\\
\hline
\multirow{2}{*}{Gaussian}&$G=2$ selected & 0\% \\
&Average ARI (std.\ dev.) & $0.4988$ ($0.06052$) \\
\hline
\end{tabular*} 
\end{table}

The five-component Gaussian solution depicted in Figure~\ref{fig:gaussEasy} is typical of 21 of the cases observed in this simulation. Particularly striking here is that the components cannot be combined to give the correct cluster memberships. Instead, there is a clear `noise' group comprising points that do not fit neatly within any of the other four components.
Although one might argue that this problem is obvious by inspection in two dimensions, and might be resolved in some \textit{post hoc} fashion, it would be virtually impossible to detect or resolve in all but very low dimensional examples.

\subsection{Old Faithful Geyser Data}\label{sec:OldFaith}
\begin{figure*}[htb!]
\centering%
\includegraphics[width=6in,height=2.5in]{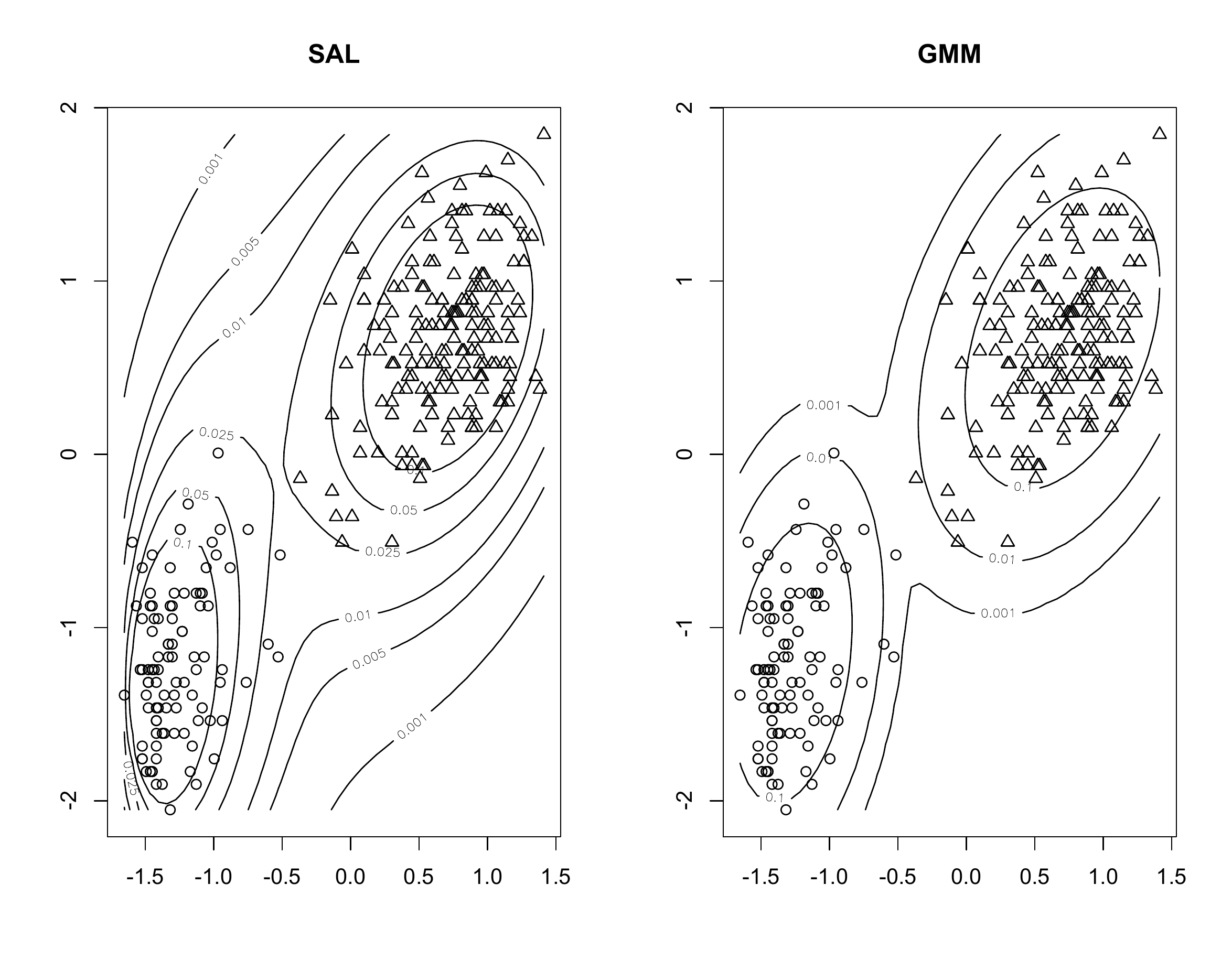}
\vspace{-0.2in}
\caption{Model-based clustering results with contours for the SAL and Gaussian mixture models (GMMs) on the Old Faithful data. \label{GeysClust}}
\end{figure*}

The famous Old Faithful geyser data, which are available as {\tt faithful} in {\sf R} \citep{R12}, comprise a two-variable data set measuring the waiting time between eruptions and the duration of 272 eruptions of the Old Faithful geyser in Yellowstone National Park. These data are well-known as an example of skewness and they have been used many times to illustrate approaches to analyzing skewed data \citep[e.g.,][]{ali10}.
Our mixture of SAL distributions and mixtures of Gaussian distributions were fitted to the geyser data. There are no `true' classifications for these data but both approaches selected sensible groups with identical classifications. The associated contour plots (Figure~\ref{GeysClust}) illustrate that the fit to the data is better for the mixture of SAL distributions. Note that applying a mixture of skew-$t$ distributions to these data will also give nicely fitting contours \citep[cf.][]{vrbik12}. However, as discussed in Section~\ref{sec:XXX}, the mixture of SAL distributions is less computationally cumbersome.



\subsection{Yeast Data}\label{sec:Yeast}
\subsubsection{The Data}
\begin{figure}[!htb]
\centering%
\vspace{-0.2in}
\includegraphics[height=2.65in]{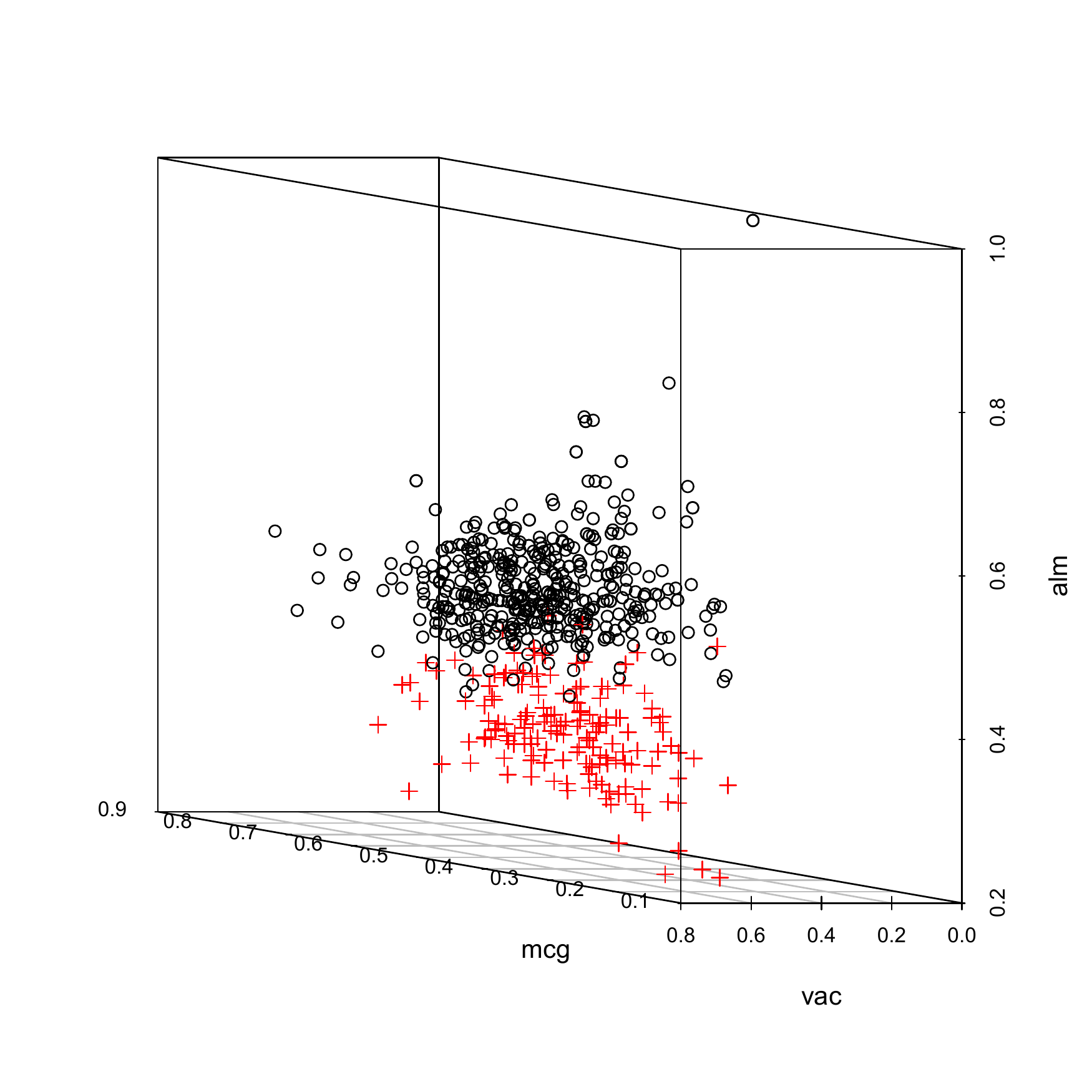}
\vspace{-0.2in}
\caption{The yeast data with the CYT and ME3 location sites highlighted.}\label{Yeast3DCol}
\end{figure}
The yeast data, which are available from the UCI machine learning repository, contain cellular localization sites of 1,484 proteins. The development of these data, as well as classification results using a `rule-based expert system', are discussed by \cite{nakai91,nakai92}. For illustration, we consider three variables: McGeoch's method for signal sequence recognition ({\tt mcg}), the score of the ALOM membrane spanning region prediction program ({\tt alm}), and the score of discriminant analysis of the amino acid content of vacuolar and extracellular proteins ({\tt vac}). The goal of our cluster analysis is to distinguish between the two localization sites, CYT (cytosolic or cytoskeletal) and ME3 (membrane protein, no N-terminal signal), cf.\ Figure~\ref{Yeast3DCol}.

\subsubsection{Model-Based Clustering Results}
\begin{figure}[!htb]
\centering%
\vspace{-0.2in}
~\hspace{-0.3in}\includegraphics[height=2.45in]{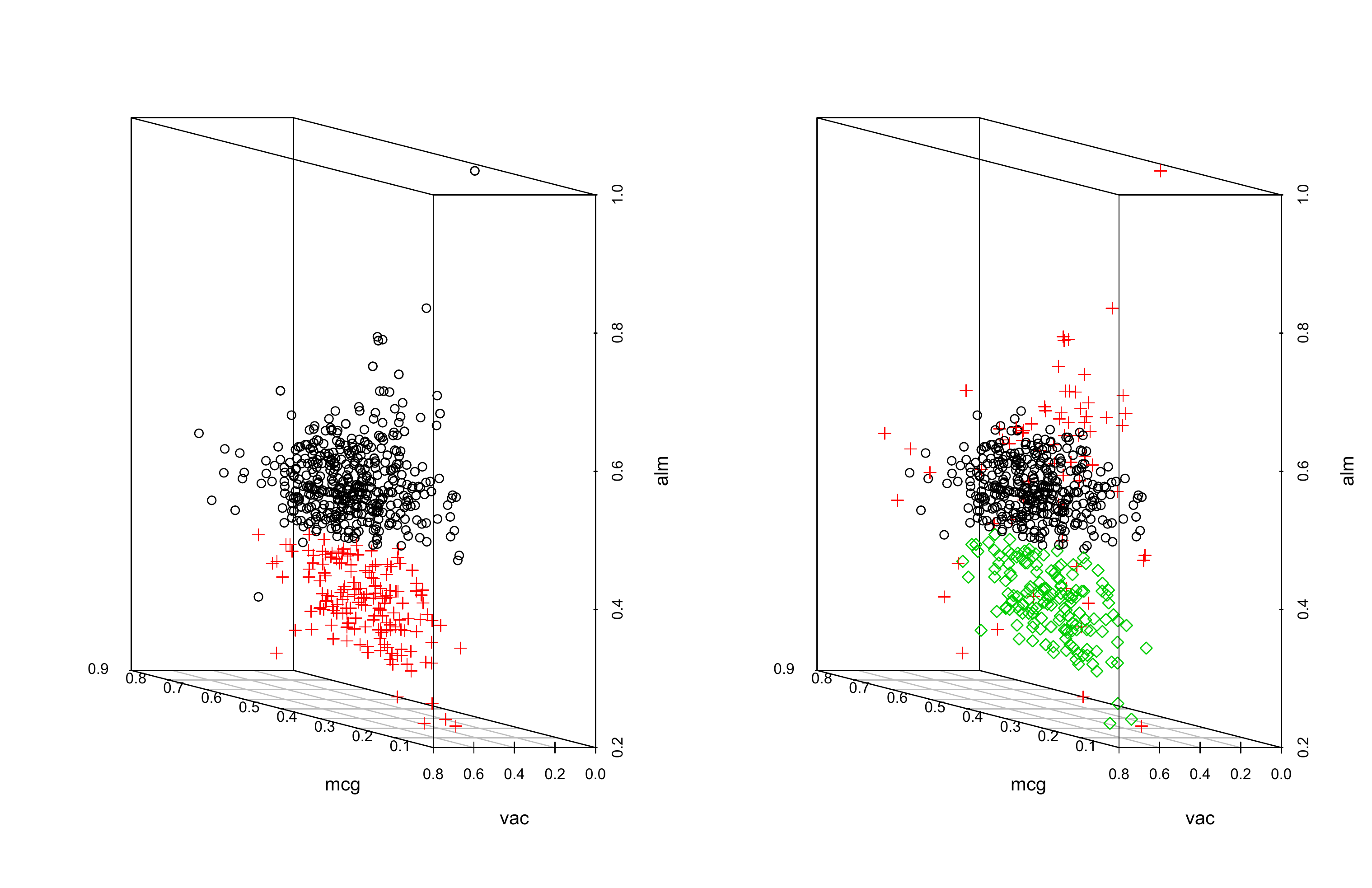}
\vspace{-0.2in}
\caption{Clustering results for the SAL and Gaussian mixture models on the yeast data; the colours highlight the predicted component memberships.\label{YeastClust}}
\end{figure}

The SAL and Gaussian mixture models were fitted to these data for $G=1,\dots,5$ components and the best fitting model was chosen using the ICL. 
The chosen SAL mixture model had two components (ICL=$-5173.133$) and the chosen Gaussian mixture model had three components (ICL=$-5161.878$). The classification performance of the SAL mixture model (ARI=$0.81$) is superior to that of the Gaussian mixture model (ARI=$0.56$), as illustrated in Table~\ref{ClustYeast} and Figure~\ref{YeastClust}. Again, this superiority cannot be negated by merging Gaussian components.%
\begin{table}[ht]
\centering%
\caption{Clustering results for the best SAL and Gaussian mixture model, as chosen by ICL, for the yeast data. The localization sites are cross-tabulated against our predicted classifications (A, B, C) in each case.}\label{ClustYeast}
\begin{tabular*}{0.48\textwidth}{@{\extracolsep{\fill}}ccccccc}
\hline
& \multicolumn{2}{c}{SAL} && \multicolumn{3}{c}{GMM} \\
\cline{2-3}\cline{5-7}
& A & B && A & B & C\\
\hline
CYT & $448$ & $15$ && $379$ & $12$ & $72$\\
ME3 & $14$ & $149$ && $13$ & $11$ & $139$\\
\hline
\end{tabular*} 
\end{table}

Now, one may argue that the Gaussian mixture model would perform better under a different model selection criterion. Therefore, we also investigated the Gaussian mixture model with $G=2$ (Table~\ref{2GYeast}). Surprisingly, this Gaussian mixture model gives very poor clustering performance, producing classifications that are worse than would be expected by guessing (i.e., ARI=$-0.088<0$). 
\begin{table}[ht]
\centering
\caption{Classification results for the two component Gaussian mixture model on the yeast data. The localization sites are cross-tabulated against our predicted classifications (A, B).}\label{2GYeast}
\begin{tabular*}{0.48\textwidth}{@{\extracolsep{\fill}}lcc}
\hline
& A & B \\
\hline
CYT & $106$ & $357$ \\
ME3 & $1$ & $162$ \\
\hline
\end{tabular*} 
\end{table}

\subsubsection{Model-Based Classification Results}
To compare model-based classification within the SAL and Gaussian mixture modelling frameworks, we analyze the yeast data with 70\% of the group memberships taken to be known. We set $G=H=2$ and each model was fitted using 25 different random 70/30 partitions of the data. The aggregate classification results (Table~\ref{ClassYeast}) and ARIs ($0.86$ and $-0.080$, respectively) indicate that the SAL mixture models outperform their Gaussian counterparts by some margin.%
\begin{table}[ht]
\centering
\caption{Aggregate classification results for the SAL and Gaussian mixture models for the yeast data with 70\% of the labels taken as known. The localization sites are cross-tabulated against our predicted classifications (A, B) for the observations with unknown labels in each case.\label{ClassYeast}}
\vspace{.1in}
\begin{tabular*}{0.48\textwidth}{@{\extracolsep{\fill}}cccccc}\hline
& \multicolumn{2}{c}{SAL} && \multicolumn{2}{c}{GMM} \\
\cline{2-3}\cline{5-6}
& A & B && A & B \\
\hline
CYT & $3403$ & $71$ && $964$ & $2502$\\
ME3 & $87$ & $1139$ && $29$ & $1205$\\
\hline
\end{tabular*} 
\end{table}

The poor performance of the Gaussian mixture modelling approach here is surprising because the parameter estimates are computed with 70\% of the location sites taken as known. This result raises questions around the efficacy of the semi-supervised Gaussian model-based classification approach, as well as further reinforcing the need for more flexible non-Gaussian approaches. Once again, the poor performance of the Gaussian approach on these data cannot be mitigated by merging components.

\section{Summary and Discussion}\label{sec:summ}

A mixture of SAL distributions model was introduced and applied for both clustering and classification. 
An EM algorithm was used for parameter estimation, with starting values obtained using the deterministic annealing approach of \cite{zhou09}. 
To account for both parsimony and entropy, the ICL is used to select the number of mixture components. 
Our model-based clustering approach was illustrated on both real and simulated data. 

In the simulation, data were generated from a SAL mixture model with two components that were very well separated. The SAL mixtures gave near-perfect results on these data whereas the Gaussian mixture models consistently overestimated the number of components. Furthermore, it is particularly notable that the overestimation of the number of components by the Gaussian mixture models could not be resolved by merging components. 

We also analyzed two real data sets. For the Old Faithful data, we considered only model-based clustering with the SAL and Gaussian mixture models and both gave the same predicted group memberships. However, a contour plot revealed that the SAL mixture model captured the shape of the data far more effectively than its Gaussian counterpart. The yeast protein location data presented a much more difficult clustering problem, and so we also used these data to illustrate model-based classification. 
For model-based clustering, the chosen SAL model gave very good clustering performance ($G=2$, ARI=$0.81$) and outperformed its Gaussian counterpart ($G=3$, ARI=$0.56$). Furthermore, when we forced $G=2$ components, the Gaussian mixture modelling approach gave worse clustering performance than would be expect by guessing (ARI$<0$). In the model-based classification applications, we set $G=H=2$ and considered 25 random subsets of  the data for which 70\% of the locations were taken as known. The SAL mixtures again gave excellent performance (ARI=0.86) but the Gaussian mixtures had negative ARIs; it is surprising that the Gaussian approach gave very poor classification performance in the semi-supervised case when 70\% of the locations were taken as known. This result calls into question the efficacy of the Gaussian model-based classification approach. Again, as with the simulation study, the poor performance of Gaussian mixtures on the yeast data could not be mitigated by merging components.

A decade on from the landmark paper of \cite{fraley02a}, we have put forth a case for substantial departure from the Gaussian model-based clustering paradigm. Unlike the skew-normal and skew-$t$ approaches, which are perhaps less substantial departures, our approach is elegant and computationally straightforward. This paper reinforces the fact that the literature is moving away from Gaussian approaches with some alacrity but it should not be taken as being pejorative of Gaussian mixture models. Gaussian mixture models remain a highly effective method for clustering and classifying certain types of data, and will forever be the midwife that introduced mixture model-based clustering. Gaussian mixtures can, however, give misleading results and, as we have illustrated, one cannot rely on rectifying poor performance by merging components. 

As mentioned in Section~\ref{estimatemu}, a better solution to the issue of infinite log-likelihood values in our EM algorithm is the subject of ongoing work.
Future work will focus on the introduction of parsimony into these models by decomposing the covariance structure in the same way as described by \cite{celeux95} for the Gaussian mixture model ($\mSigma_g = \lambda_g\mathbf{D}_g'\mathbf{A}_g\mathbf{D}_g$). Analysis of very high-dimensional data with our SAL mixtures will be explored along the lines of work by \cite{mcnicholas08,mcnicholas10d} and \cite{baek10}, who focused on extensions of mixtures of factor analyzers. Handling noise will be considered, both via the addition of a uniform component \cite[cf.][]{banfield93} and through the trimmed clustering procedures of \cite{gallegos05}. Work will also be conducted on developing a model selection technique to compare clustering results from SAL mixtures with those from other mixtures, such as mixtures of skew-$t$ distributions. 

\section*{Acknowledgements}
This work was supported by an Ontario Graduate Scholarship, the University Research Chair in Computational Statistics, and an Early Researcher Award from the Ontario Ministry of Research and Innovation.

\bibliographystyle{chicago}
\bibliography{laplace}

\end{document}